# Review: Progress on 2D topological insulators and potential applications in electronic devices

Yanhui Hou(侯延辉), Teng Zhang (张腾), Jiatao Sun(孙家涛), Liwei Liu(刘立巍)†, Yugui Yao(姚裕贵), Yeliang Wang(王业亮)

MIIT Key Laboratory for Low-Dimensional Quantum Structure and Devices, Beijing Institute of Technology, Beijing, China

Two-dimensional topological insulators (2DTI) have attracted increasing attention during the past few years. New 2DTI with increasing larger spin-orbit coupling (SOC) gaps have been predicted by theoretical calculations and some of them have been synthesized experimentally. In this review, the 2DTI, ranging from single element graphene-like materials to bi-elemental TMDs and to multi-elemental materials, with different thicknesses, structures and phases, have been summarized and discussed. The topological properties (especially the quantum spin Hall effect and Dirac fermion feature) and potential applications have been summarized. This review also points out the challenge and opportunities for future 2DTI study, especially on the device applications based on the topological properties.

## 1. Introduction

Topological insulator is one of the most emergent research areas in condensed matter physics, especially for the novel physical phenomena. As a classic example, graphene was successfully exfoliated from graphite bulk to single-atom-thickness layer in 2004[1] and was predicted to be a two-dimensional topological insulators (2DTI), dubbed as Quantum spin Hall insulator (QSHI), by C. L. Kane and E. J. Mele in 2005.[2, 3] Since then, the exploration of 2DTI has gained increasing interest due to its scientific importance as a novel quantum state and related applications.[4-6]

---

† Corresponding author. E-mail: liwei.liu@bit.edu.cn

2DTI hold several significant advantages due to the unique electronic structures. First, electronic structures of 2DTI have gaps in the bulk and gapless edge states (Fig. 1 (a)). Such edge states have the spin-momentum locking property (Fig. 1 (b)), which can suppress electronic backscattering against nonmagnetic impurity due to the time-reversal (TR) symmetry. Second, the edge states of 2D TI have the Dirac fermions feature, that is, a linear dispersion relationship in the energy space (Fig. 1 (a)) which results in high mobility carriers.[7, 8] Compared with three-dimensional (3D) materials, the 2D materials have quantum confinement in the vertical direction, resulting a distinctly different band structure.[9] Moreover, 2DTI have significant advantages in the tunability of the energy band, valley and spin properties as they can be easily manipulated by external fields like magnetic,[10] electric[11], optic,[12, 13] and strain field.[14] Therefore, the 2DTI is very promising for dissipationless electronic devices, quantum computation, ultrafast response opto-electrical devices *et al*.[7, 8]

The prediction of quantum spin Hall effect (QSHE) in graphene has inspired exploration of other 2DTI materials with stronger SOC strength. In 2006, B. A. Bernevig *et al*. theoretically predicted that QSHE can be achieved in HgTe/CdTe quantum well[15], which was soon realized experimental in 2007 by Molenkamp's group[16]. Since then, there has a wave of scientific exploring research on 2DTI. Later, people quickly spread it to other quantum well systems theoretically[17-19] and experimentally[20-22]. The initial discovery and subsequent development of 2DTI in semiconductor quantum well systems have played a pivotal role, and have been discussed in previous papers like that in Reviews of Modern Physics[7-9].This review will focus on the brilliant progress of 2DTI in the recent years, especially on the relationship between the geometric structures and the topological properties, as well as the related applications. In the end, we will give the perspective on the challenge and opportunities for future study on 2DTI.

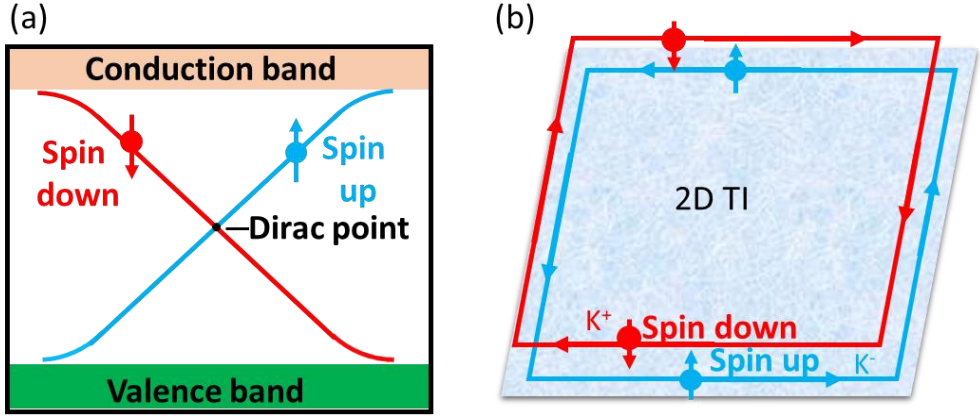

**Fig. 1** (a) Band structure of two-dimensional topological insulators (2DTI), which has the gap in the bulk denoted by the separation between conduction band from valence band, and gapless states at the edge denoted by the blue and red curves. The blue and red branch carries spin up and down current, respectively (b) Spin-momentum locking in the edge states of 2DTI in real space.

## 2.  Topological properties and application of novel 2D materials

During the last decade, new members of 2DTI with increasing SOC gap have been predicted theoretically and prepared experimentally, such as silicene, germanene, bismuthene, $Bi_4Br_4$, BiX/SbX (X=Cl, Br, I) and 1T' TMD. The typical theoretically predicted values of SOC gaps are listed in Table 1. We further analyze the development of each of these materials, the related topological properties and the potential applications.

**Table 1**: The values of SOC gap in different 2DTI.

| 2DTI | graphene | silicene | germanene | bismuthene | BiX/SbX | $Bi_4Br_4$ | 1T'-TMD |
|---|---|---|---|---|---|---|---|
| SOC gap (meV) | $0.8 \times 10^{-3}$ [2] | 1.55 [23] | 23.9 [23] | 470 [24] | 1080 [25] | 180 [26] | 150 [27] |

## 2.1.  Silicene and Germanene

Silicene, the silicon analogue of graphene, was proposed to host quantum spin Hall effect by Liu C. C. *et al.*[28] in the year of 2011. Compared to graphene, silicene has a stronger SOC gap due to the heavier atomic mass and the buckled geometry. According to the DFT calculations, silicene has a SOC gap of 1.55 meV (Fig. 2(a)) with helical gapless edge states. Although silicene has a buckled structure, the band structure also has a massless Dirac fermion behavior with a Fermi velocity of ~$10^6$ m/s, as fast as that of graphene. Due to the compatibility with the current established silicon-based semiconductor industry, it seems silicene can be a good material for field effect transistor (FET). However, the bandgap of silicene is still small (the corresponding QSH critical temperature is ~18 K) and may hinder its applications in the electronic devices. Various methods have been tested to induce a larger gap (a topological trivial band gap) in silicene layer, such as adsorption or intercalation of metal atoms[29-31], hydrogenation[32, 33] and oxidization[34]. Similar to graphene nanoribbon, 1D silicene nanoribbon, made of silicene by cutting, has a tunable bandgap up to about 0.4 eV, which was predicted by DFT calculations [5, 35] and realized in experiments[36, 37].

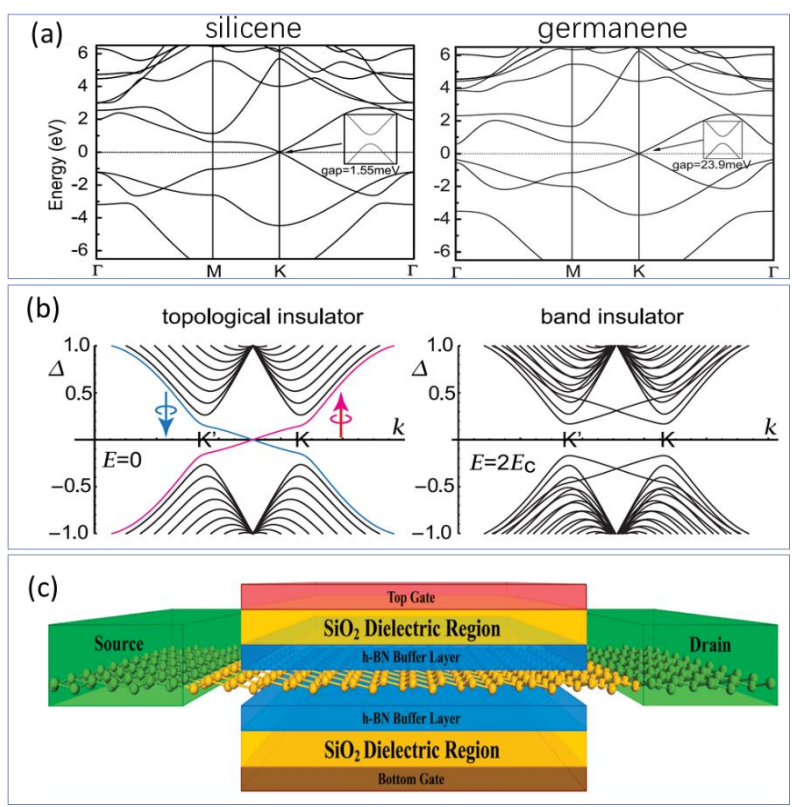

**Fig. 2** (a) Calculated electronic band structures of buckled silicene and germanene. (b) Band structures of silicene or germanene under different vertical electric field E, which turns a topological insulator to a band insulator (c) Schematic model of dual-gated silicene FET. (a) Reprinted with permission from Ref. [23]. Copyright 2011, American Physical Society. (b) adopted with permission from Ref. [38]. (c) Reprinted with permission from Ref. [39]. Copyright 2012, American Chemical Society.

Apart from the above chemical or structural modification methods, which may affect the fast response of topological property-based devices, another method by applying a vertical electric field has been proposed.[38] The vertical electric field can break the symmetry of the two sublattices of silicene and induce a topological phase transition between the QSH edge state (ON) and trivial insulator state (OFF), as shown in Fig. 2(b). Utilizing this feature, silicene could be a candidate for a new type of quantum mechanical switch, so-called topological insulator field effect transistor (TI-FETs), as shown in Fig. 2(c).[39] Thus silicene provides a flexible platform for low-power consumption and unconventional electronic devices based-on topological property.

In Periodic Table of Elements, another Group-IVA element, germanium also is predicted to form a single layer stable structure, named as germanene. Germanene is also 2DTI with a low-buckled honeycomb structure similar to silicene, but a much larger spin-orbit gap of 23.9 meV[23] was predicted (Fig. 2(a) right panel), corresponding to the QSH Tc of 276 K. Thus QSHE, as well as germanene-based TI-FETs, are expected to be observed experimentally.

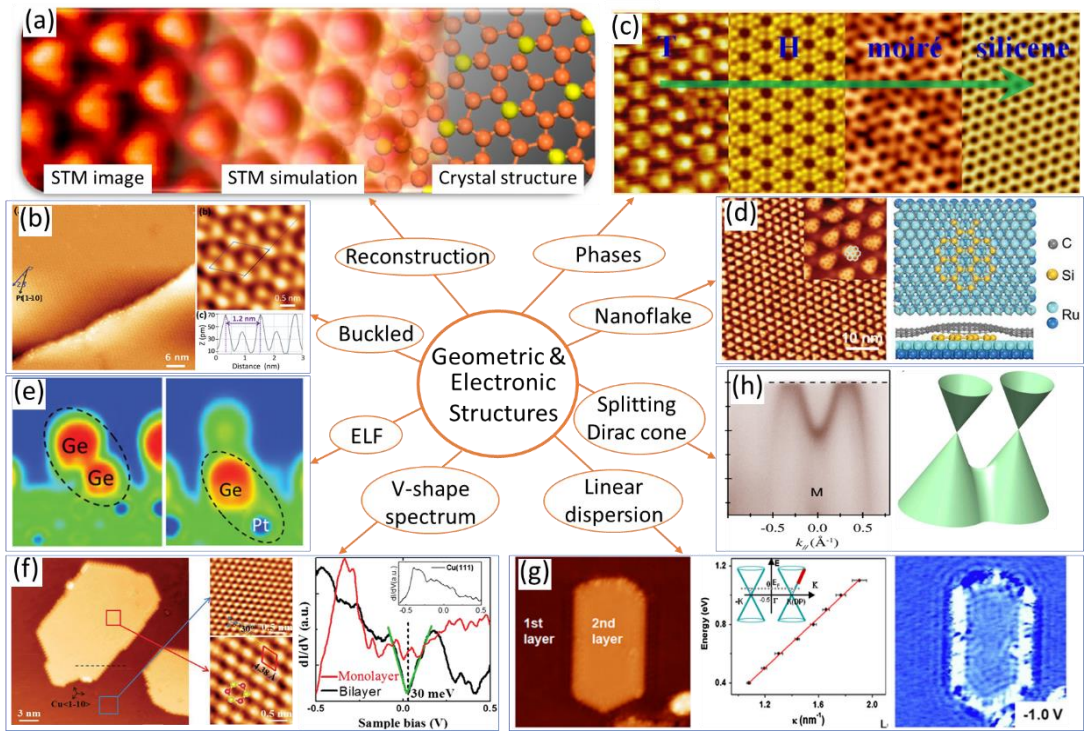

**Fig. 3** Geometric and electronic structures of silicene and germanene. (a) STM image, theoretical simulated image and crystal structure of silicene on Ir(111) from left to right. (b) STM image of buckled germanene on Pt(111). (c) T and H phase of silicene on Ag(111) substrate. (d) STM images

of silicene nanoflake intercalated between graphene and Ru(0001). (e) Calculated electronic localization function (ELF), showing the covalent interaction existing between adjacent germanium atoms and electrostatic interaction between germanium atom and substrate. (f) dI/dV on monolayer (red) and bilayer (black) germanene. (g) dI/dV mapping on the 2nd layer silicene island which can derive the linear dispersion relationship of Dirac Fermion. (h) Splitting of Dirac cone observed by angle resolved photoemission (ARPES) on 3×3-silicene on Ag(111). (a) Reprinted with permission from Ref. [40]. Copyright 2014, WILEY-VCH Verlag GmbH & Co. KGaA, Weinheim. (c) Reprinted with permission from Ref. [41]. Copyright 2012, American Chemical Society. (d) Reprinted with permission from Ref. [42]. Copyright 2018, WILEY-VCH Verlag GmbH & Co. KGaA, Weinheim. (f) Reprinted with permission from Ref. [43]. Copyright 2017, WILEY-VCH Verlag GmbH & Co. KGaA, Weinheim. (g) Reprinted with permission from Ref. [44]. Copyright 2012, American Physical Society. (h) Reprinted with permission from Ref. [45]. Copyright 2019, American Physical Society.

Shortly after the theoretic prediction of silicene and germanene, there have been great progresses on the experimentally findings, both on characterization of geometric structures and electronic structures. The high-quality samples have been synthesized by molecular beam epitaxy (MBE) method on different substrates.[40, 46-48] For example, High-quality silicene with a ($\sqrt{3}\times\sqrt{3}$) silicene/($\sqrt{7}\times\sqrt{7}$) Ir(111) superstructure on Ir(111)(Fig. 3(a)),[46] and silicene with both T and H phases on Ag(111) were independently reported (Fig. 3(c)).[41] Besides, germanene with buckled configuration was reported on Pt(111) and Cu(111) substrates (Figs. 3(b) and (f)).[40, 43]Calculated electron localization function (ELF) showed the covalent interaction existing between adjacent germanium atoms, while the electrostatic interaction between germanium atom and substrate, thus supporting the 2D nature of germanene lattice (Fig. 3(e)).

Because of the easy oxidation of silicene in air, usage of silicene for device application become a challenge. To solve this problem, Gao's group intercalated silicon between graphene and Ru(0001) to protect silicene layer by covering of graphene layer. Depending on the silicon coverage, silicene nanoflake, single layer and multilayer layer silicene were prepared underneath graphene, as shown in Fig. 3(d).[42] As for device, silicene-based FET were reported in 2015, which demonstrated theoretical expectations of its ambipolar Dirac charge transport and showed a high carrier mobility of $\sim 100$ cm$^2$ V$^{-1}$ s$^{-1}$ at room-temperature.[49]

Regarding the electronic structures, studies mainly focus on the linear dispersion feature of Dirac Fermions. For example, In 2017, high-quality bilayer germanene islands on Cu(111) was reported, which revealed a V-shape dI/dV spectrum, indicating a signature of Dirac-type electrons (Fig. 3(f)).[43] By dI/dV mapping on bilayer silicene islands at different bias voltages around the Fermi level, Wu's group have successfully described the linear dispersion relationship of the edge states, which is a hallmark of Dirac fermions.[44] More recently, they reported a new physical phenomenon- the splitting of Dirac cones measured by ARPES technique, in the system of 3×3-silicene on Ag(111), ascribed to external periodic potentials from the substrate-overlayer interaction (Figs. 3(g) and (h)). [45]

## 2.2. Antimonene

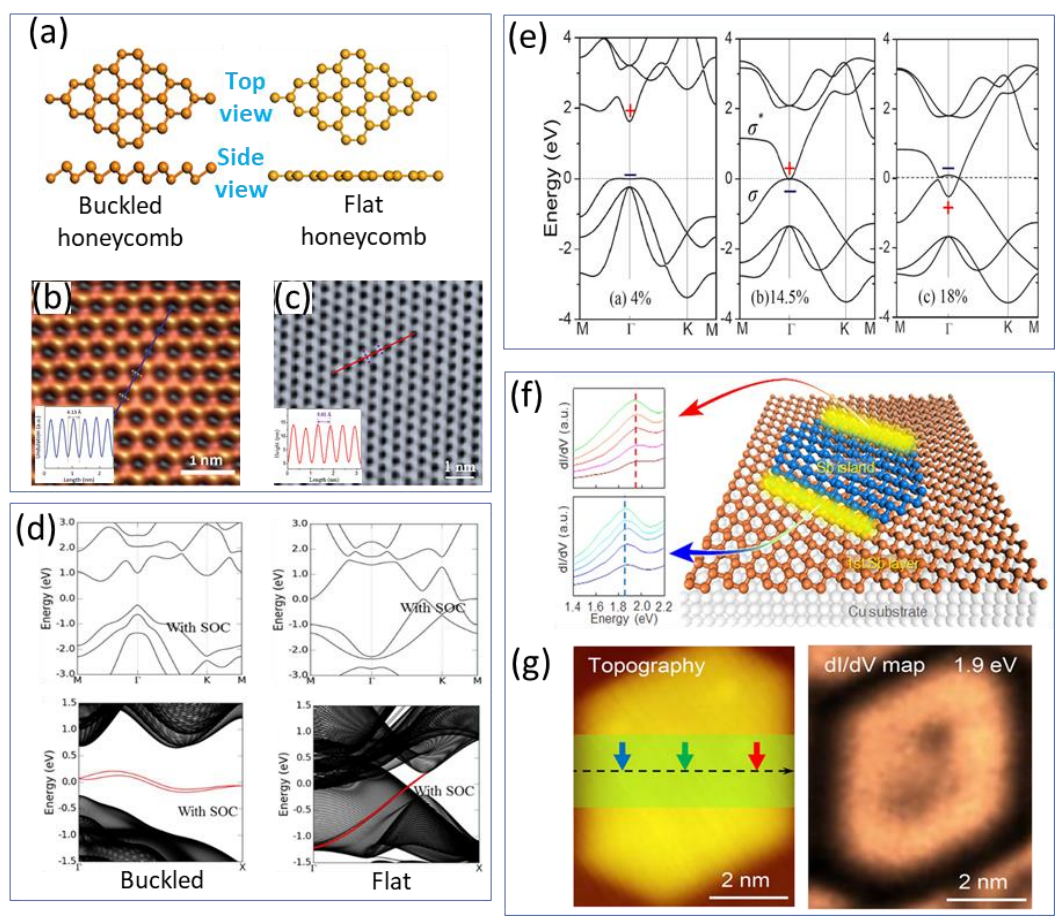

**Fig. 4** Geometric and electronic structures of antimony monolayer. (a) Atomic structures of the buckled and flat honeycombs of antimony. (b), (c) Atomic resolution STM image of the buckled antimonene monolayer (BAM) and flat antimonene monolayer (FAM). Inset: line profile corresponding to the line in panel. (d) Band structures and the edge states of antimony monolayer. FAM

has a topologically nontrivial edge state and BAM has a topologically trivial edge state. Red lines denote the edge states. (e) Inverted band of antimony monolayer induced by increasing strain field. (f) Edge states of antimonene island on top of ML antimonene. (g) STM image and dI/dV mapping of the antimonene island. (a, c, d) Reprinted with permission from Ref. [50]. Copyright 2018, American Chemical Society. (b) Reprinted with permission from Ref. [51]. Copyright 2017, WILEY-VCH Verlag GmbH & Co. KGaA, Weinheim. (e) Reprinted with permission from Ref. [14]. Copyright 2015, Nature Publishing Group. (f, g) Reprinted with permission from Ref. [52]. Copyright 2019, American Chemical Society.

Antimonene, a 2D honeycomb lattice made of antimony atoms, is another attractive mono-elemental candidate for 2D topological material.[53-55] Two structures of antimonene have been found: buckled honeycomb and flat one (Fig. 4 (a)), which were successfully fabricated on $PdTe_2$ and Ag(111) substrates by MBE, respectively.[50, 51] Figures 4(b) and 4(c) show atomic-resolution STM images of buckled antimonene monolayer (BAM) and flat antimonene monolayer (FAM). The calculated electron localization function (ELF) of Sb-Sb pair are higher than that between Sb and the substrate, which proves that antimonene of both structures are in 2D nature with minor interaction with the substrate. Band calculations show that FAM is a topological semimetal with topologically non-trivial edge state, while BAM have a topologically trivial edge state (Fig. 4(d)). A theoretical work reporting that the bulk bandgap of buckled antimonene can be tuned under different tensile, and above tensile of 14.5%, the bands are inverted and creating a topological gap (Fig. 4(e)).[14] Thus, the flat and buckling structures provide a feasible way to regulate the electronic properties of antimonene and extends related nanodevice applications. Very recently, the antimonene nanoislands in the second layer with topologically nontrivial one-dimensional edge states has been reported (Figs. 4(f) and (g)).[52] This work not only provides hallmark to topologically nontrivial states in the BAM but it also opens a novel route for tailoring the topological edge states by external strain of the supporting substrate.

## 2.3. Bismuthene, $Bi_4Br_4$ and BiX/SbX (X=H，F，Cl，Br)

Following the trend of searching for large SOC 2DTI, monolayer or few-layer materials with the high-Z element Bismuth (Z=83) are explored.[56-58] In 2014,

Yazdani's group reported one-dimensional topological edge states of bismuth bilayers on top of bulk bismuth.[59] Importantly, one type of edge showed spectra with a density of states that has inverse-square-root singularity inherent to a 1D states. Moreover, by visualizing the quantum interference of edge-mode quasi-particles in the confined geometries, they demonstrated the coherent propagation along the edge with properties consistent with strong suppression of backscattering. In 2017, Reis *et al.*[60] reported that the bismuthene on a SiC as a candidate for a high temperature quantum spin Hall material. Figures 5(a) and 5(b) present the geometric structure and band structure of bismuthene on a SiC, respectively. The line dI/dV across the step edge reveals a large bulk gap of ~0.8 eV with local metallic edge state (Fig. 5(c)). As the bismuthene was prepared on a semiconductor SiC substrate, this work pushed forward the future application for realizing various device proposals with topological edge modes.

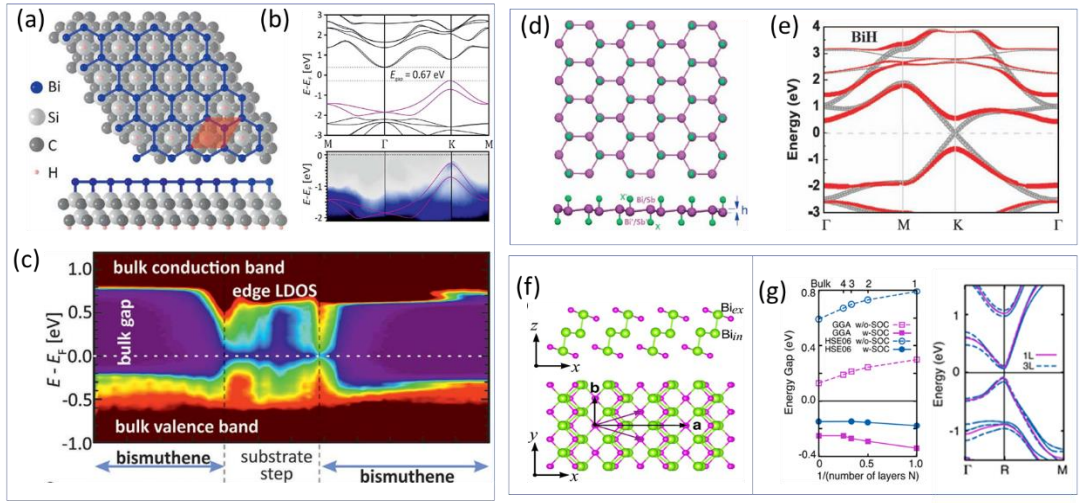

**Fig. 5** Geometric and electronic structures of bismuthene and BiX layers. (a) Sketch of a bismuthene layer placed on the threefold-symmetric SiC(0001) substrate. (b) DFT calculated electronic band structures and ARPES experimental electronic band dispersion of single layer bismuthene. (c) Spatially resolved dI/dV data across the step, the dI/dV signal of the in-gap states/peaks at the edges of bismuthene film. (d) Atomic structure of monolayer BiX/SbX (X=H，F，Cl，Br), by top view and side view. (e) Electronic band structure of BiH and BiF with large gap. Red line means SOC is considered, but the gray line is not. (f) Side view and top view of the crystal structure of single-layer $Bi_4Br_4$. (g) The edge state and spin polarization of different layers of $Bi_4Br_4$. The layer dependence of energy gap of few-layer $Bi_4Br_4$ (left panel), and band structures of single-layer and triple-layer $Bi_4Br_4$ (right panel). (a-c) Reprinted with permission from Ref. [60]. Copyright 2017, American Association for the Advancement of Science. (d, e) Reprinted with permission from Ref. [61]. Copyright 2014, Nature Publishing Group. (f, g) Reprinted with permission from Ref. [62].

Besides the 2D TI materials made of single element, the search for 2D TI with a larger gap is also expanded to bi-element materials like BiX/SbX (X= H, F, Cl and Br) and $Bi_4Br_4$. Because Bi and Sb are well known for their strong SOC that can drive and stabilize the topological states, it is intuitive to search for large-band-gap QSH insulators based on Bi/Sb-related materials. In 2014, Yao's group predicted a group of 2DTI BiX/SbX (X=H, F, Cl and Br) monolayers with extraordinarily large bulk gaps based on first-principles calculations[61] and tight-binding calculations[25], respectively. Figures 4(d) and 4(e) showed the unit cell and the typical band structures of BiH and BiF. 2D BiX has a honeycomb structure like silicon hydride (silane), with a threefold rotational symmetry like that in graphene. The side view exhibits the nearest neighbor Bi-X distances and buckling heights, a low-buckled configuration is more stable for the BiF, BiCl and BiBr monolayers.

Interestingly, the giant-gaps are due to the strong SOC related to the $p_x$ and $p_y$ orbitals of the Bi/Sb atoms, in contrast to $p_z$ orbital in graphene and silicene. The topological characteristic of BiX/SbX monolayers is confirmed by the calculated nontrivial $Z_2$ index. The bulk band gap of greater than 1.0 eV in BiH and BiF monolayers is probably the largest bulk band gap of all the reported TIs predicted theoretically. Moreover, BiX/SbX monolayers in vertical electrical field become a quantum valley Hall insulator, exhibiting valley-selective circular dichroism.

While many 2DTI with large SOC gap have been proposed and realized in the experiment, the lack of appropriate insulating substrates is another crucial issue. Single-layer $Bi_4Br_4$ may break this obstruction because its bulk crystal is an insulator. [62] Single-layer $Bi_4Br_4$ can be obtained by mechanical exfoliated from its bulk material (see Fig. 5(f) for the layered crystal structure), and was recently demonstrated to be a QSH insulator with sizable gap (0.18 eV)[26]. The energy gap and band structures of few-layer $Bi_4Br_4$ do not change much with increasing layers (Fig. 5(g)). Furthermore, at the boundary of multilayer $Bi_4Br_4$, the topological edge states stemming from different single-layers are weakly coupled, and can be fully decoupled by constructing a stair-stepped edge. Thus the decoupled topological edge states are very suitable for multi-channel dissipationless transport.

## 2.4. 1T'-TMD materials

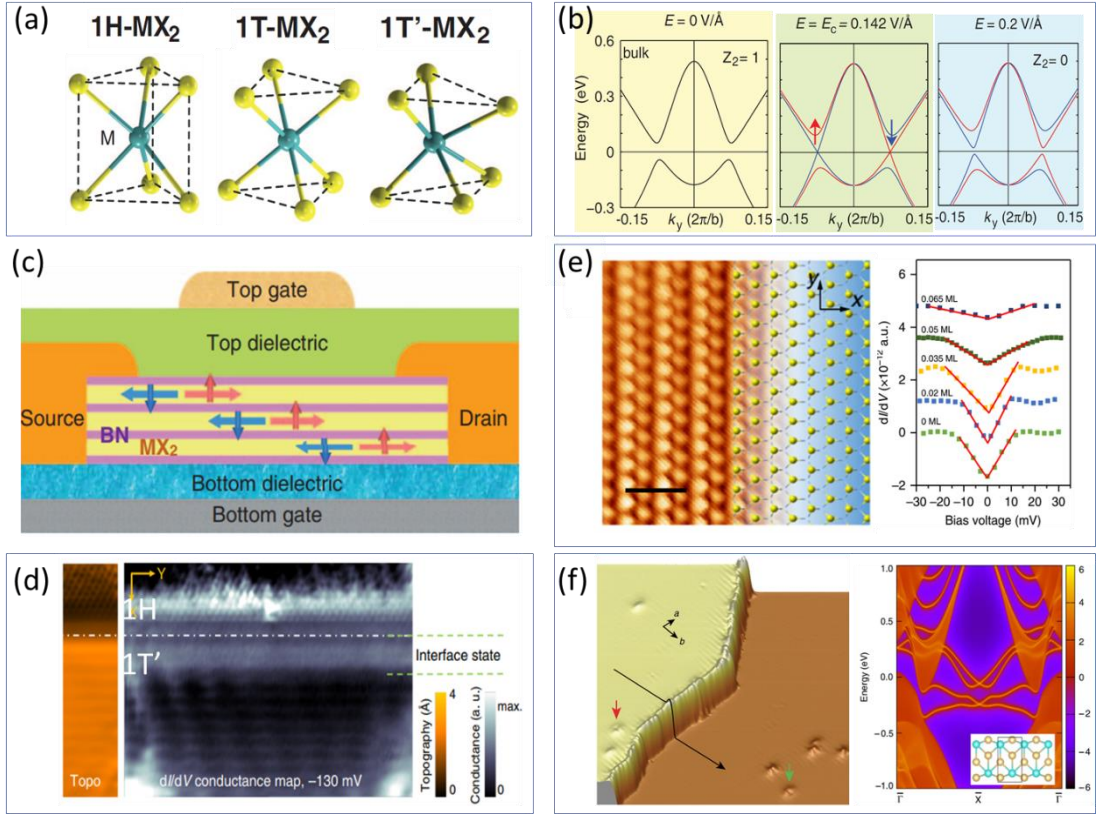

**Fig. 6** Geometric and electronic structures of TMDs. (a) Three typical atomic configurations of transition metal dichalcogenides (TMDs, MX₂). (b) Band inversion of 1T'-MX₂ induced by vertical electric fields. (c) A vdW-FET by stacking several layers of 1T'-MX₂ and 2D dielectric layers (h-BN). (d) Experimental dI/dV map taken at the 1T'−1H interface. (e) Atomic resolution STM image of the single-layer 1T'-WTe₂ surface and dI/dV fitted with a linear function near the Fermi energy. (f) Pseudo 3D image of WTe₂ showing a step edge and the calculated topological edge state of the monolayer 1T'-WTe₂ bulk electronic structure. (a-c) Reprinted with permission from Ref. [63]. Copyright 2014, American Association for the Advancement of Science. (d) Reprinted with permission from Ref. [64]. Copyright 2018, Nature Publishing Group. (e) Reprinted with permission from Ref. [65]. Copyright 2018, Nature Publishing Group. (f) Reprinted with permission from Ref. [66]. Copyright 2017, Nature Publishing Group.

The 2D layered transition metal chalcogenides (TMDs, MX₂) (M = transition metal, X = chalcogen) have received increasing attention in recent years. As shown by geometric schematics in Fig. 6(a), TMDs have a variety of phases such as 2H, 1T, 1T'. Among them, 2H and 1T phases have been studied in 2D semiconductor systems such as 2H-MoS₂, 1T-PtSe₂. Recently the 1T' phase is gaining increasing attention due to the type-II Weyl semimetal in the bulk WTe₂. While the properties of ML 1T'-MX₂ has not been unveiled until in 2014, Qian *et al.*[63] studied 1T'-MX₂ (with M =(tungsten or molybdenum) and X = (tellurium, selenium, or sulfur), and predicted that it changed from a type-II Weyl semimetal in the bulk form to a class of large-gap QSH materials in the monolayer form. The structural distortion from 1T to T' doubles the lateral period and causes an intrinsic band inversion, and SOC interaction further opens a gap of 0.08 eV, rendering the topological properties. A vertical electric field breaks inversion symmetry and introduces a strong Rashba splitting of the doubly degenerate bands near the fundamental gap Eg. As the field increases, Eg first decreases to zero at a critical field strength of 0.142 V/Å and then reopens (Fig. 6(b)). This gap-closing transition induces a topology change to a trivial phase. Similar to the $Bi_4Br_4$, the 2D 1T'-$MX_2$ have shown possibility in constructing van-der Waals devices like Topological FET (T-FET) by stacking multilayer together to increase the conducting channels (Fig. 6(c)).

This theoretic prediction has inspired a wave of experimental study on the preparation and topological study on the monolayer 1T'-$MX_2$. In 2017, Crommie's group[67] reported the study of electronic structure of the monolayer 1T'-$WTe_2$ on BLG/SiC(0001) by a combination of different techniques like ARPES, DFT and STM. They observed topological band inversion, band gap opening and the edge states that were consistent with expectations for a QSH insulator. In addition, this group found similar QSH behavior in the system of ML 1T'-$WSe_2$ (Fig. 6(d)).[64] Using the 1D interfaces between the topologically non-trivial 1T' and 1H phases, they confirmed the predicted penetration depth (2 nm) of one-dimensional interface states into the 2D bulk of a quantum spin Hall insulator (QSHI). Li *et al.*[65] used high-resolution quasiparticle interference (QPI)-STS/STM to study the monolayer 1T'-$WTe_2$ also on BLG/SiC(0001), and found that it had a semimetal bulk band structure without a full SOC-induced band gap. Importantly, there is a Coulomb gap induced by electron–electron interactions at the $E_F$ (Fig. 6(e)). Not only monolayer $WTe_2$ but also the bulk step edges can exhibit the QSHE. In 2018, Peng *et al.*[66] reported 1D step edge state of $WTe_2$ observed by STM (Fig. 6(f)). Importantly, the edge state shows significant robustness against edge imperfections and is unaffected by the presence of the substrate, which provides evidence for the topological origin.

## 2.5. Magnetic Topological Insulators

2D topological insulators were originally proposed in time-reversal invariant systems, however, the interplay between magnetic order and nontrivial topology can provide fertile ground to explore new physics and applications such as spintronics with low energy consumption, dissipationless topological electronics and topological quantum computation. Recently, van der Waals layered $AB_2C_4$ materials (such as $MnBi_2Te_4$ and $SnSb_2Te_4$) have been predicted to host intriguing topological properties. For example: In 2019, Yong Xu's group[68] predicted $MnBi_2Te_4$ to show intralayer ferromagnetic and interlayer antiferromagnetic, as well as rich topological quantum states such as a type II magnetic Weyl semimetal, intrinsic axion insulators and QAH insulators in even- and odd-layer films, respectively. Using ARPES, Hong Ding's group[69] and E. V. Chulkov's group[70] independently reported the observation of surface state in $MnBi_2Te_4$, respectively, and provided clear evidence for nontrivial topology of the antiferromagnetic TIs. Very recently, various of exciting properties such as Quantum anomalous Hall effect[71], robust axion insulator and Chern insulator phases[72] were observed in $MnBi_2Te_4$ thin flakes by method of quantum transport. These $MnBi_2Te_4$ family materials can be expected to integrate with other 2D materials to form van der Waals heterostructures and provide great opportunities for exploring exotic topological phenomena.

## 2.6. Quasi-2D Topological Materials and the optoelectronic coupling

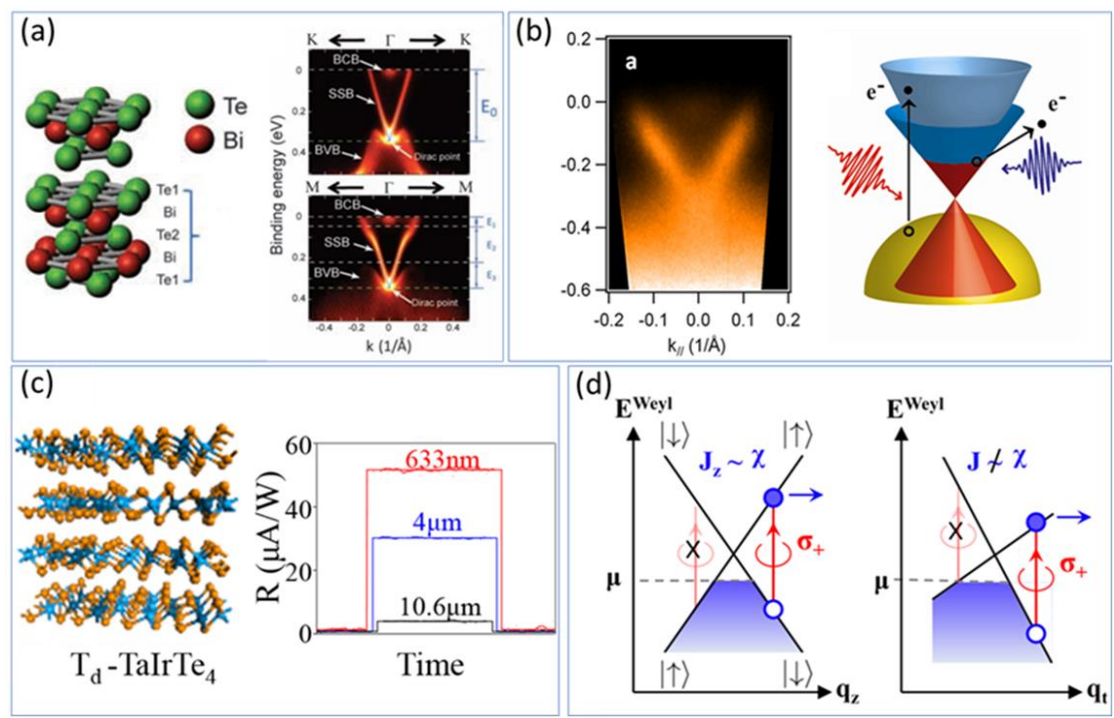

**Fig. 7** Quasi-2D topological materials and the optoelectronic coupling. (a) Crystal structure and ARPES images of Bi$_2$Te$_3$. (b) ARPES image and schematic of infrared laser pump excites electrons from the valence band (orange) into the conduction band of Bi$_2$Te$_3$. (c) Crystal structure of the Type-II WSM TaIrTe$_4$ and Broadband photoresponse of TaIrTe$_4$ photodetector. (d) Schematics of photocurrent generations in type I (left panel) and type II (right panel) Weyl systems. (a) Reprinted from Ref. [73]. Copyright 2009, American Association for the Advancement of Science. (b) Reprinted with permission from Ref. [74]. Copyright 2012, American Chemical Society. (c) Reprinted with permission from Ref. [75]. Copyright 2018, American Chemical Society. (d) Reprinted with permission from Ref. [76]. Copyright 2017, American Physical Society.

While the optoelectronic coupling of 2DTI utilizing the topological properties have been limited, there are many examples of 3DTI which have been explored for this purpose. For example: In 2009, Shen's group reported the investigation of the surface state of Bi$_2$Te$_3$ with angle-resolved photoemission spectroscopy, proving its 3DTI nature with a single Dirac cone on the surface (Fig. 7(a)). The large bulk gap of Bi$_2$Te$_3$ is promising for potential applications in high-temperature spintronics. Hajlaoui *et al.* [74] used time-resolved ARPES to study the transient electronic dynamics of Bi$_2$Te$_3$, and provided a direct real-time visualization of the transient carrier population of both the surface states and the bulk conduction band (Fig. 7(b)). McIver *et al.* [77] found that by illuminating Bi$_2$Se$_3$ with circularly polarized light, a helicity-dependent photocurrent which originated from topological Dirac fermions can be generated. Moreover, a photocurrent which is controlled by the linear polarization of light may also have a topological surface state origin. $Bi_2Se_3$ and $Bi_2Te_3$ in these devices are layered nanoscale materials and can be regarded as stacked 2D materials, thus we call them "quasi-2D" materials. The opto-electronic coupling of 3DTI also extended to 3D Weyl semimetal. Weyl semimetals were transformed from Dirac materials by breaking the space inversion or time reversal symmetry. The break of symmetry splits the degeneracy of Dirac point into two Weyl points with opposite chiralities, the electronic excitations of which can be described by the Weyl equation.[78, 79] In 2018, Sun's group reported the realization of a broadband self-powered photodetector based on $TaIrTe_4$ (Fig. 7(c)). Compared with the graphene and $Cd_3As_2$, $TaIrTe_4$ improved the response time from ps to us order. In 2019, Sun's group[80] further reported the realization of nonlinear optical correspondence on the type-II Weyl semimetal $TaIrTe_4$ sheet. Figure 7(d) illustrates the mechanism of generation of photocurrent in Weyl systems. 3D Weyl semimetals can generically support significant photocurrents due to the inversion symmetry breaking and also finite tilts of the Weyl spectra for type-II Weyl semimetal. Compared with the traditional photocurrent based on PN junction, 3D Weyl semimetals need neither applied electric field nor control of the depletion layer interface, and has a faster response on fs order, which will be very suitable for the next generation of ultrafast response photoelectric devices.

## 3. Summary and perspectives

As mentioned above, there have been enormous progress on the study of 2DTI, from discovering of new members (including new phases) with increasing larger SOC gaps to exploring novel properties. For the application using topological properties, such as QSH effect and the Dirac fermion feature, many obstacles have been removed. For example: the largest SOC gap is over 1 eV, corresponding to a temperature much higher than room temperature so that application based on the QSH effect is possible. 2DTI can be prepared not only on metal substrates, but also on semiconductors and insulators, thus the interaction with substrates can be tuned and back gate can be applied for the device purpose. With van-der Waals layered materials like ML $Bi_4Br_4$ and 1T' TMD, the number of conducting channels ($e^2/h$ per edge, where $e$ is the elementary charge and $h$ is Planck's constant) can be increased by stacking several ML materials together, resulting in a larger signal-to noise ratio.

However, there are still some challenges and open questions for future developments of 2DTI: (a) The intrinsic topological characteristics of 2D systems need to be further understood. (b) More 2D topological systems combining several advantages (with high topological critical temperatures, on different substrates and can be stacked to increase conducting channels) still needs to be discovered. (c) Efficient methods of fast on/off switching are still lacking. Device preparation to tune the 2DTI with magnetic, electric, optic and strain field need to be further explored. (d) Although a lot of theoretical papers predicting the application of 2DTI, till now the practical realization of 2DTI has been limited. A lot of the 2DTI are made into devices utilizing the semiconducting or valley properties, instead of the topological properties. Meanwhile, the applications of topological systems mainly focus on the 3D systems, for example, 3DTI $Bi_{14}Rh_3I_9$ which hosts the helical edge states and offers the opportunity to design spin filters[81], while the 2D system is much fewer. Although these materials are 3DTI, the extension to 2DTI can be expected. Thus, there is a great future to expand the area of 2DTI and make devices based on the topological properties into practice.

## Acknowledgments

L.W.L., Y.L.W. and T. Z. thank the Beijing Natural Science Foundation (Nos. Z190006, 4192054), National Natural Science Foundation of China (Nos. 61971035, 61901038, 61725107), Strategic Priority Research Program of the Chinese Academy of Sciences (XDB30000000), and Beijing Institute of Technology Research Fund Program for Young Scholars (3050011181814).